# Third Order Modal Exceptional Degeneracy in Waveguides with *Glide-Time* Symmetry

Farshad Yazdi, Tarek Mealy, Alireza Nikzamir, Robert Marosi, and Filippo Capolino
Department of Electrical Engineering and Computer Science
University of California, Irvine, CA 92697 USA

*Abstract*—The dispersion of a three-way waveguide is engineered to exhibit exceptional modal characteristics. Two coupled waveguides with *Parity-Time* (*PT*) symmetry have been previously demonstrated to exhibit second order exceptional points of degeneracy (EPDs). In this work, we introduce and investigate a particular class of EPDs, applicable from radio frequency to optical wavelengths, whereby three coupled waveguides satisfy *Glide-Time* (*GT*) symmetry to exhibit a third order modal degeneracy with a real-valued wavenumber. *GT* symmetry involves glide symmetry of lossless/gainless components of the waveguide in addition to changing the sign of passive/active elements while applying a glide symmetry operation. This *GT*-symmetry condition allows three Floquet–Bloch eigenmodes of the structure to coalesce to a real-valued wavenumber at a single frequency, in addition of having one branch of the dispersion diagram with a purely real wavenumber. The proposed scheme may have applications including but not limited to distributed amplifiers, radiating arrays, and sensors, from radio frequency to optics.

## I. Introduction

We propose and investigate a periodic three-way electromagnetic waveguide with a *glide-time (GT)*-symmetric topology that exhibits a distinguished class of degeneracy conditions based on the coalescence of three degenerate modes with a real wavenumber. The concept presented in this paper is based on applying concepts inspired by *PT* symmetry [1,2] to a glide-symmetric waveguide [3]. We call this combination *GT* symmetry. It is different from *PT* symmetry since the waveguide does not possess parity symmetry. In this paper we show that a waveguide with GT symmetry, i.e., with a balanced condition of gain and loss, possesses an exceptional point of degeneracy (EPD) of order three, with a real-valued wavenumber.

Exceptional degeneracies of order 2, 3, and 4 of eigenmodes in periodic media have been previously investigated in [4–9], demonstrating the existence of unique features associated with modal degeneracies; even though they did not name them "exceptional points", they provided the math and physics associated to such degeneracy points. Exceptional points and their perturbation have been studied previously in more general terms [10–13] (note that the term exceptional point was already mentioned in the 1966 book of Kato [12], Ch. 2.). These degeneracies are not just in the eigenvalues but also in the polarization states (eigenvectors). The concept of EPD associated with the coalescence of modes is relatively recent in the study of active devices. The recent interest in this class of degeneracies was mainly motivated by their relevance in the study of *Parity-Time-* (*PT*-) symmetric systems in physics [1,2,14–24].

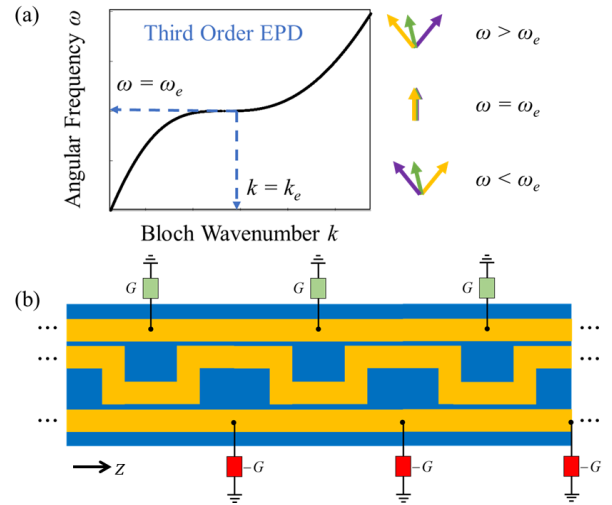

Fig. 1. (a) An example of dispersion relation of the mode with purely real wavenumber in an infinitely long periodic waveguide made of three coupled waveguides with loss and gain satisfying *GT* symmetry. The third order EPD occurs at the angular frequency $\omega_e$ with real-valued Floquet-Bloch wavenumber $k_e$, where three eigenvectors (schematically represented by three vectors) coalesce. (b) As an example, the 3-way periodic waveguide with third order EPD is made of three coupled microstrips over a grounded substrate (in blue) with periodic gain and loss, shifted by half a period. The structure in (b) can be seen in a more general way as two transmission lines coupled through a third serpentine transmission line. The 3-way periodic waveguide supports three modes in each longitudinal direction.

The *GT*-symmetric waveguide in this paper is implemented by adding balanced gain and loss (it can be radiation loss due to antenna radiation), to a glide-symmetric waveguide. A periodic waveguide is said to possess glide (*G*) symmetry if it remains invariant under the glide operation, consisting of a translation by half of the geometrical period, *d*, followed by a reflection in the so-called glide plane [3,25–32]. We define *GT* symmetry as regular glide symmetry of lossless and gainless components of the waveguide, in addition to changing the sign of passive/active elements while applying such glide symmetry operation. In other words, it is a combination of *PT* symmetry and glide symmetry.

Previously, different kinds of EPDs have been found in lossless guiding structures in [4,5,19,33,34]. In particular, an EPD of order three in a lossless waveguide, also called stationary inflection point (SIP) has been demonstrated in [33,35–37], whereas an EPD of order four, namely the degenerate band edge (DBE), has been demonstrated in multimode lossless waveguides [6,38–43].

At radio frequency (RF), the SIP was experimentally demonstrated in a three-way waveguide made of three coupled microstrips in [37]. The occurrence of the DBE has also been experimentally demonstrated at RF in [40,41,44]. In [45], the authors experimentally demonstrated a split band edge, which is a degeneracy closely related to the DBE, in a metallic circular waveguide loaded with anisotropic scatterers.

On the other hand, EPDs of order 2, 3, and 4 have been demonstrated theoretically and experimentally in [2,46–52] and [16,44,53,54], respectively, by proper balancing of the loss/gain, using the concept of *PT* and anti-*PT* symmetry.

The general subject of this paper is the investigation of third order modal degeneracy in a three-way waveguide with balanced loss and gain, satisfying *GT* symmetry. For waveguides made of three coupled transmission lines (i.e., three ways) like the one we consider in this work, the allowed orders of EPD are $2^{nd}$, $3^{rd}$, $4^{th}$, and $6^{th}$. The $3^{rd}$ order EPD is the only one that does not have a stopband above or below the EPD frequency (it is the only odd order) and has a group velocity that does not change sign above and below the EPD frequency. This makes the $3^{rd}$ order EPD beneficial for amplifier applications. (For an amplifier application of an SIP, i.e., a $3^{rd}$ order EPD without gain and loss, see [55].)

In the vicinity of third-order EPDs, the dispersion diagram of eigenmodes in a periodic waveguide satisfies $(\omega - \omega_e) \propto (k - k_e)^3$, where $\omega_e$ is the angular frequency at which three modes coalesce and $k_e$ is the real-valued Bloch wavenumber at the degeneracy point. Note that $k_e \neq \pi/d$, meaning that the EPD will not occur at the edge, or middle, of the Brillouin zone. An illustration of an ideal dispersion relation exhibiting a third order EPD is shown in Fig. 1(a), where only the real branch of the $\omega - k$ dispersion diagram (where $k$ is the Bloch-wavenumber and $\omega$ is the angular frequency) is shown. This kind of degeneracy obtained in a lossless waveguide has been named SIP. Here, instead, we investigate the occurrence of analogous third order EPDs in *GT*-symmetric waveguides, i.e., where both gain and losses are present. The periodic set of losses in the *GT*-symmetric waveguide in Fig. 1, represent the radiation resistances of an array of antennas.

The fundamental concept offered here is potentially useful for a variety of applications. Indeed, the use of the DBE has been proposed already for low threshold oscillators with a stable oscillation frequency [56–58]. Recently the DBE oscillator has been experimentally demonstrated in [59]. Oscillators based on EPD with balanced loss and gain have been proposed in [47,48,60,61] that are, in principle, able to radiate high power.

The SIP application has been proposed for delay lines [36]. High efficiency, high gain amplifiers based on SIP have also been proposed in [55] based on the concept of *three-mode synchronization*, in traveling wave tubes.

The third order EPD studied here can be applied to the case of distributed amplifiers interleaved with an array of antennas for high power radiation, since, in principle, the EPD can be designed with large gain balanced with large radiation loss.

The paper is organized as follows: in Section II, we introduce and discuss the two kinds of unit cell structures for the three-way waveguide, where the transfer matrix of the unit cell is modeled using coupled transmission lines (CTLs). The modal dispersion of the periodic structure is investigated where we demonstrate the existence of third order EPDs in the dispersion diagram for a few designs. We also provide a thorough analysis of the power distribution for the semi-infinite structure as well the engineering of the dispersion diagram to have different characteristics by tuning the parameters of the unit cell. Section III is dedicated to the finite length studies of the periodic structure with proper terminations where we study the resonance behavior and stability through the *S*-parameters of the three-way waveguide. We also investigate the power performance of the finite-length structure for a distributed radiating amplifier application and its important characteristic aspects such as stability analysis and radiating and load power gains. Throughout this paper, we implicitly assume that the time-dependence is in the form of $e^{j\omega t}$.

## II. THREE-WAY COUPLED WAVEGUIDE WITH *GT* SYMMETRY

We define *GT* symmetry as the combination of two operators: the G- glide symmetry, and the *T*- time reversal symmetry operators. In the modeling of an optics/electromagnetic system, the time reversal operators *T* makes the imaginary unit $j \to -j$, hence when applied to a refractive index, it implies that $n(x,z) \to n^*(-x,z)$, i.e., loss goes into gain and vice versa. The glide symmetry operators *G* makes a translation by half of the geometrical period, *d*, followed by a reflection in *x*. In terms of refractive index, it implies that $n(x,z) \to n(-x, z+d/2)$, The glide symmetry is considered a higher symmetry. The combined *GT* operator leads to $n(x,z) \to n^*(-x, z+d/2)$.

In the following we investigate a three-way waveguide that satisfies these properties. However, we focus on a metal-dielectric structure with lumped loss and gain, which is described in more details in the next section.

The goal of this paper is to show that a structure that satisfies *GT* symmetry has a third order EPD with real-valued wavenumber. The study of the spectrum of the *GT* operator is left to future investigations.

### A. Unit Cell Design of the Coupled Serpentine Waveguide with Gain and Loss

We consider two distinct periodic waveguide geometries in microstrip technology based on the three-way CTLs with unit cells as shown in Fig. 2. The designs are modeled by two uniform transmission lines that are coupled through a third serpentine-shaped transmission line in the middle, similar to the structure in [37]. In this paper, we have altered the structure by adding balanced gain and loss. This is implemented using a set of periodic lumped line-to-ground admittances on the first (top) microstrip with a conductance of $-G$ (gain) and another set of periodic lumped line-to-ground admittances in the third (bottom) microstrip with the conductance of $+G$ (loss, or radiation loss) to achieve a *GT*-symmetric design for the three-way microstrip structure. We find the degeneracy condition of order 3 by selecting proper periodic loss and gain values. The third order EPD is analogous to the SIP, which can be found in passive, lossless three-way waveguides. However, the presence of lumped gain and loss elements makes the system more complicated. In terms of applications, the periodic gain provides amplification and losses may represent discrete radiating elements (e.g., antennas). Therefore, this scheme can be viewed either as a distributed radiating amplifier or as a structure that may radiate and oscillate (i.e., lasing) at the same time. Moreover, the degeneracy may bring advantages in terms of low noise, enhanced coherency among the radiating elements, etc.

We provide two potential implementations of such a *GT*-symmetric structure: in Fig. 2(a), the discrete elements of gain/loss are located at the uncoupled sections of the CTLs whereas in Fig. 2(b) they are located at the coupled sections. We provide in Appendix A the design parameters for both structures in Fig. 2. We assume that the conductance, $G$, has a pure real value representing either loss or gain in the structure. The three-way CTL supports three modes in each longitudinal direction. Thus, the structure can exhibit a third order EPD by tuning the microstrip geometry and admittances. In designing the unit cell to attain the EPD, for the sake of simplicity, it is assumed that all the transmission lines have the same width $w$, same separation distance between the coupled lines, $s$, and the length of each unit cell is set to $d$. We used a substrate with a relative dielectric constant of 2.2, no loss tangent (tan($\delta$)=0), and height of $h_s$ = 1.575 mm. Also, the microstrip and ground plane metal layers were assumed to be lossless. To achieve the degeneracy condition at the desired frequency ($f_e$ = 2 GHz), we fixed values for some of the dimensions including $w$ = 5 mm for the line widths (corresponding to lines with $Z_0$ = 50 Ohm characteristic impedance, when uncoupled) and $s$ = 0.5 mm for the distance between the lines. We then tune other dimensions such as the length of the unit cell, $d$, the height of the serpentine section, $h$, and the value of the conductance, $G$, to search for the third order degeneracy at the desired frequency. The optimization we have done to find a third order EPD is based on tuning the prementioned parameters to minimize the coalescence parameter associated with three eigenmodes in the system, as will be discussed later.

### B. Transfer Matrix Formalism

We use a three-CTL transfer matrix formalism to construct the total transfer matrix for a single unit cell, in analogy to what was done in [37,44,48]. We will also use this transfer matrix in our analysis of the finite-length periodic structure composed of cascaded unit cells to model and investigate the various aspects of the modal degeneracy under study. The details of the transfer matrix formalism are provided in Appendix B. Other related matrix-based approaches have previously been used to analyze systems under *PT* and broken *PT* symmetry regimes in works such as [48,62–66].

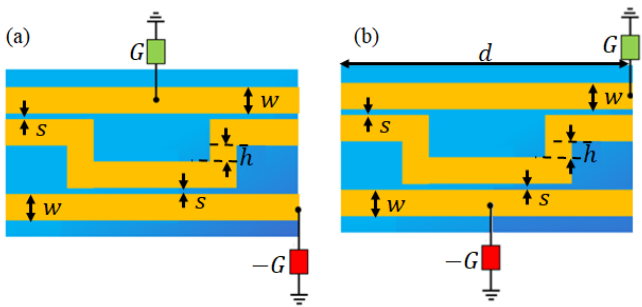

Fig. 2. (a) Unit cell of the 3-way periodic microstrip structure that exhibits a third order degeneracy for Case *A* (also used in Secs. III and IV). The structure is composed of two uniform transmission lines coupled through a third serpentine transmission line in the middle, and two shunt conductances, gain and passive (e.g., a radiation resistance) added to the uncoupled sections with real positive and negative values as shown. (b) Unit cell of an alternative design (Case *B* and Case *C*) of the 3-way periodic microstrip structure where the shunt conductances (gain and loss) are added to the coupled sections instead.

In the investigation of the EPDs' properties through transfer matrix and eigenvalues for a 6-port system, it is convenient to define the position-dependent state-vector in the form

$$\mathbf{\Psi}(z) = [V_1, \ Z_0 I_1, \ V_2, \ Z_0 I_2, \ V_3, \ Z_0 I_3]^T, \quad (1)$$

where voltages and currents are evaluated at $z$ along the three-way CTLs. The state vector describes the spatial evolution of the eigenmodes as they propagate through the structure. A *transfer matrix*, $\underline{\mathbf{T}}(z_2, z_1)$, is used, which uniquely relates the state vector $\mathbf{\Psi}(z)$ between two points in the structure such that

$$\mathbf{\Psi}(z_2) = \underline{\mathbf{T}}(z_2, z_1)\mathbf{\Psi}(z_1), \quad (2)$$

where we use the forward transfer matrix notation with $z_2 > z_1$ along the $z$ axis. The 6×6 transfer matrix $\underline{\mathbf{T}}_U$ of a unit cell shown in Fig. 2(a), is then defined as $\mathbf{\Psi}(z+d) = \underline{\mathbf{T}}_U \mathbf{\Psi}(z)$ and is expressed and calculated in terms of the geometric and electrical parameters of the unit cell using formulas found in Appendix B. Accordingly, the unit cell transfer matrix for the waveguide in Fig. 2(a) is obtained by cascading the transfer matrices of each segment of the unit cell

$$\underline{\mathbf{T}}_U = \underline{\mathbf{T}}_{-G} \underline{\mathbf{T}}_A \underline{\mathbf{T}}_C \underline{\mathbf{T}}_B \underline{\mathbf{T}}_{+G} \underline{\mathbf{T}}_B \underline{\mathbf{T}}_C \underline{\mathbf{T}}_A. \quad (3)$$

The expression for unit cell transfer matrix for Fig. 2(b) is presented in Appendix B. For an infinitely long stack of CTL unit cells, a pseudo-periodic solution for the state vector $\mathbf{\Psi}(z)$ exists in the Bloch form and the transfer matrix $\underline{\mathbf{T}}_U$ translates the state vector across a unit cell as the eigenvalue equation of

$$\underline{\mathbf{T}}_U \mathbf{\Psi}(z) = e^{-jkd}\mathbf{\Psi}(z), \quad (4)$$

where $k$ is the complex-valued Bloch wavenumber. The eigenvalues of the transfer matrix and hence the Bloch wavenumber are obtained as solutions of the characteristic equation

$$\mathrm{Det}\left(\underline{\mathbf{T}}_U - \zeta \underline{\mathbf{I}}\right) = 0, \quad (5)$$

in which we define $\underline{\mathbf{I}}$ to be the $6\times 6$ identity matrix. For the CTL with three lines ($6\times 6$ transfer matrix) discussed in this study, six eigenvalues, $\zeta_i = e^{-jk_i d}$, with $i=1,2,...,6$, of the $\underline{\mathbf{T}}_U$ matrix are calculated from equation (5).

Note that, because of periodicity, each eigenvalue corresponds to an infinite set of wavenumbers $k_i + n2\pi/d$, with $n = 0, \pm 1, \pm 2,...$, called Floquet harmonics. In the following, we show the dispersion diagrams with wavenumbers in the range $0 < \mathrm{Re}(k) < 2\pi/d$ that we refer to as the fundamental Brillouin zone.

Because of the reciprocity of the system, the transfer matrix satisfies $\mathrm{Det}(\underline{\mathbf{T}}_U) = 1$. Consequently, if $\zeta$ is an eigenvalue of the system then $\zeta^{-1}$ is another eigenvalue. Therefore, the modes supported by the structure have wavenumbers $k_1, k_2, k_3, -k_1, -k_2$ and $-k_3$. At the third order EPD studied in this paper, three eigenvalues coalesce at $k_e$ while the other three coalesce at $-k_e$. Moreover, at the EPD, the transfer matrix $\underline{\mathbf{T}}_U$ cannot be diagonalized because the three eigenvectors of (4) associated with each $k_e$ and $-k_e$ wavenumber, coalesce, as discussed in [37,42]. The coalescence of three eigenvectors is a necessary and sufficient condition for a third order EPD to occur. This means that the existence of an EPD can be found by checking the coalescence of three eigenvectors. This is the technique implemented in this paper to find the EPD conditions while maintaining *GT* symmetry. At the EPD, only two polarizations states, $\mathbf{\Psi}_{e1}$ and $\mathbf{\Psi}_{e2}$, are the eigenvectors of the system. This implies that the geometric multiplicity of each degenerate eigenvalue is equal to 1 while its algebraic multiplicity is equal to 3, hence the transfer matrix $\underline{\mathbf{T}}_U$ is not diagonalizable and it is similar to a matrix containing two Jordan blocks of dimensions 3 × 3, as explained in details in [42]. At the EPD, the transfer matrix $\underline{\underline{T}}_U$ is represented as

$$\underline{\underline{T}}_U = \underline{\underline{V}} \begin{bmatrix} \underline{\underline{\Lambda}}_{J,1} & \underline{\underline{0}} \\ \underline{\underline{0}} & \underline{\underline{\Lambda}}_{J,2} \end{bmatrix} \underline{\underline{V}}^{-1}, \quad (6)$$

where $\underline{\underline{\Lambda}}_{J,1}$ and $\underline{\underline{\Lambda}}_{J,2}$ are two Jordan blocks

$$\underline{\underline{\Lambda}}_{J,1} = \begin{bmatrix} \zeta_e & 1 & 0 \\ 0 & \zeta_e & 1 \\ 0 & 0 & \zeta_e \end{bmatrix}, \quad \underline{\underline{\Lambda}}_{J,2} = \begin{bmatrix} \zeta_e^{-1} & 1 & 0 \\ 0 & \zeta_e^{-1} & 1 \\ 0 & 0 & \zeta_e^{-1} \end{bmatrix}, \quad (7)$$

and the similarity transformation matrix $\underline{\underline{V}}$ is composed of one degenerate eigenvector and two generalized eigenvectors, associated with each of the eigenvalues $\zeta_e$ and $\zeta_e^{-1}$.

The theory explained in [42] is for a lossless three-way waveguide but there are many similarities with the waveguide in this paper which has periodic gain and loss elements. Also in this paper, we find a branch of the dispersion diagram that corresponds to a purely real wavenumber (shown in Fig. 1), while the other two branches in Figs. 3-5 represent waves with complex wavenumbers, as discussed in the next section.

### C. Dispersion Relation and Coalescence Parameter Featuring Third Order EPD

The periodic three-way microstrip in Fig. 2 can support a third order degeneracy. We design three different CTLs (Cases *A*, *B*, *C*) where the EPD occurs at an operating frequency of 2 GHz. Our unit cell designs have been determined by using the fixed parameters provided in Appendix A, such as the microstrip width, spacing between coupled microstrips, substrate dielectric properties, and substrate thickness. We then tuned the other parameters such as the length of the unit cell *d*, the "height" *h* of the serpentine sections, and the choice of lumped gain and loss conductances *G* and *–G*, respectively, to obtain EPDs at a desired frequency. Both the EPD frequency and the flatness of the dispersion curve in the vicinity of the degeneracy condition can be altered by tuning the dimensional and electrical parameters of the unit cell.

An EPD is represented by the coalescence of the eigenvalues (i.e., wavenumbers) and by the coalescence of the eigenvectors (i.e., polarization states). The coalescence of the eigenvalues is necessary to have an EPD, however, the coalescence of the eigenvector guarantees the existence of an EPD. In the following, we assess the occurrence of a third order EPD by observing the coalescence of three eigenvectors. Accordingly, we define a figure of merit to measure how close the system is to an ideal third order degeneracy condition at the frequency of interest, called Coalescence *Parameter* ($C_{EPD}$). This concept was developed in [44] for a fourth order degeneracy, and used also in [37] for an SIP; it is here analogously defined here for a third-order EPD as

$$C_{EPD} = \frac{1}{3} \sum_{\substack{m=1, n=2 \\ n>m}}^{3} |\sin(\theta_{mn})|, \quad \cos(\theta_{mn}) = \frac{\mathrm{Re}|(\boldsymbol{\Psi}_m, \boldsymbol{\Psi}_n)|}{\|\boldsymbol{\Psi}_m\|\|\boldsymbol{\Psi}_n\|}, \quad (8)$$

where $\theta_{mn}$ represents the angle between two eigenvectors $\boldsymbol{\Psi}_m$ and $\boldsymbol{\Psi}_n$ in a six-dimensional complex vector space, with norms $\|\boldsymbol{\Psi}_m\|$ and $\|\boldsymbol{\Psi}_n\|$, and $(\boldsymbol{\Psi}_m, \boldsymbol{\Psi}_n)$ is their inner product. The coalescence parameter defined in equation (8) is always positive, with small values indicating how well the eigenvectors of the structure are close to each other in the frequency range of interest. EPDs of third order occur when $C_{EPD} = 0$. Using this coalescence parameter as the error function to be minimized at the EPD frequency of interest, an optimization algorithm in MATLAB was used to select the conductance of the lumped elements, serpentine height *h*, and period of our unit cell *d* to make the device exhibit an EPD of third order.

We provide three examples of EPDs that occur in three-way microstrip waveguides as in Fig. 2, denoted as Cases *A*, *B* and *C*. These cases were each found using the optimization method discussed above.

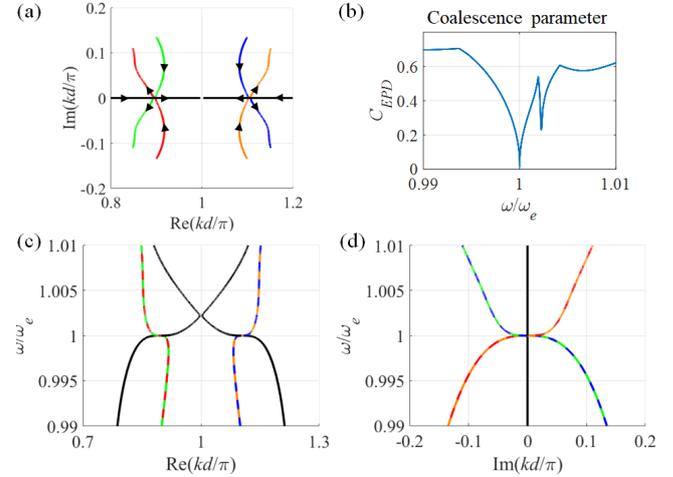

Fig. 3. Case *A*: (a) complex wavenumbers plotted in the complex *k* plane varying frequency. This plot shows the existence of the third order modal degeneracy condition and coalescence of the three modes at two different EPD real-valued wavenumbers, $k_e$ and $-k_e + 2\pi/d$. It also shows than one branch is purely real. (b) Coalescence parameter plotted versus normalized frequency in the vicinity of the EPD. (c) and (d) Typical modal dispersion diagram of the eigenmodes, showing both the real and imaginary parts of the normalized complex Floquet–Bloch wavenumber *k* versus normalized angular frequency around the EPD frequency $\omega_e$. The purely real branches are shown in solid black. Dashed-line branches represent the modes with complex wavenumber, using the same colors as in (a). Besides the lumped elements, we have assumed the three-way waveguide to be lossless for all the graphs shown in this figure.

Case *A*: In this example, the tuned unit cell parameters were found to have a conductance value of $G = 0.1398$ S (or equivalently $R = 1/G = 7.15$ Ω), serpentine height of $h = 5.35$ mm, and period of $d = 54.15$ mm. The active (gain) and passive conductances in this case are located on the unit cell as illustrated in Fig. 2(a).

Figure 3 shows the existence of third order degeneracy in the dispersion diagram and the coalescence parameter. The imaginary part of the dispersion diagram is plotted versus the real part in Fig. 3(a) where it shows the existence of third order degeneracy condition and the coalescence of the three modes at two different locations in the fundamental Brillouin zone, at $k_e$ and $-k_e + 2\pi/d$ due to reciprocity. In other words, we show an EPD in the region $0 < kd < \pi$, in the dispersion diagram of Fig 3(c). There are three coalescing branches, one (in solid black) has a purely real wavenumber with positive group velocity for frequencies around the EPD frequency as can be seen by the black curve on the left side of Fig 3(c). The second EPD is in the region $\pi < -kd + 2\pi < 2\pi$, where there are three coalescing branches. One branch (solid black) has a purely real wavenumber with negative group velocities for frequencies around the EPD. Through the rest of the paper, we consider the mode in the region $0 < k_e d < \pi$ as our EPD of the interest associated with forward waves in our dispersion diagram.

In Fig 3. (b), the Coalescence Parameter is plotted versus normalized frequency around $\omega_e$ (corresponds to 2 GHz) to demonstrate how close we are to the third order degeneracy condition in our design. And finally, in Fig. 3(c) and (d), we plot the modal dispersion diagram of the infinite structure, showing both the real and imaginary parts of the normalized complex Floquet–Bloch wavenumber $k$ versus normalized angular frequency around the designed frequency $\omega_e$ where the third order behavior is observed. We used dashed lines in Fig. 3(c) and (d) for wavenumbers that are complex valued to show different overlapping curves of real and imaginary parts. In other words, the curves with dashed lines of different colors represent two overlapping branches. We follow the same scheme in Fig. 4 and Fig. 5.

The normalized dispersion relation around the desired third order EPD can be approximated using the third order equation

$$(\omega/\omega_e - 1) \approx \zeta (kd/\pi - k_e d/\pi)^3 \quad (9)$$

where $\omega_e$ is the angular frequency at which three modes coalesce and $k_e$ is the Floquet-Bloch wavenumber at the degeneracy point. The nondimensional parameter $\zeta$ determines the flatness of the normalized dispersion at the EPD which is related to the third derivative of $d^3\omega/dk^3$ around the degeneracy point. Lower values of the flatness factor $\zeta$ mean flatter dispersion relations at the EPD which is an important factor in designing a 3rd order EPD for possible applications based on the desired characteristics and properties. For Case $A$ shown in Fig. 3 the flatness factor is calculated as $\zeta_A \approx 2.1$.

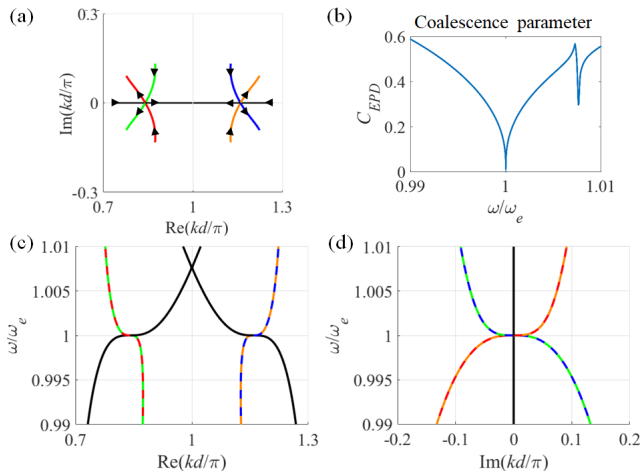

Fig. 4. Case $B$: The description is as in Fig. 3 but plots are for Case $B$.

Case $B$: Using the same optimization method, we find additional solutions which exhibit third order modal degeneracy. The unit cell design of Case $B$ differs from that of Case $A$ in that the lumped elements are positioned in the center of the coupled sections of the transmission line, as is illustrated in Fig. 2(b). For this second solution, the tuned unit-cell was found to have the conductance value of $G = 0.105$ S (or equivalently $R = 1/G = 9.5$ Ω), serpentine height of $h = 6.36$ mm, and period of $d = 46.3$ mm. Like the previous case, the dispersion diagrams and coalescence parameter are plotted in Fig. 4, where we show the existence of the EPD for the new values and discuss its modal behavior. For the case $B$ shown in Fig. 4, the flatness factor is calculated as $\zeta_B \approx 7.2$ which is higher than Case $A$, meaning a narrower dispersion diagram compared to the previous case (i.e., less flat).

Case $C$: To show the flexibility of our design we have provided a third solution that exhibits third order modal degeneracy in its dispersion diagram by again tuning the dimensions around initial values which seem appropriate for a practical design and search for a new set of parameters to achieve the third order EPD. The tuned unit-cell parameters were found to be a conductance value of $G = 0.0099$ S (or equivalently $R = 1/G = 100.55$ Ω), serpentine height of $h = 1.07$ mm, and period of $d = 48.08$ mm. As in Case $B$, the lumped elements are centered in the CTL sections, as illustrated in Fig. 2(b). The existence of the third order EPD for this case is shown in the results of the Fig. 5, where we have plotted the dispersion diagram and the coalescence parameter in a fashion similar to the previous cases. For the case $C$ shown in Fig. 5, the flatness factor is calculated as $\zeta_C \approx 188$ which is much higher than the two previous cases $A$ and B indicating a more-narrow EPD in the dispersion diagram, as can be seen from the results in Fig. 5.

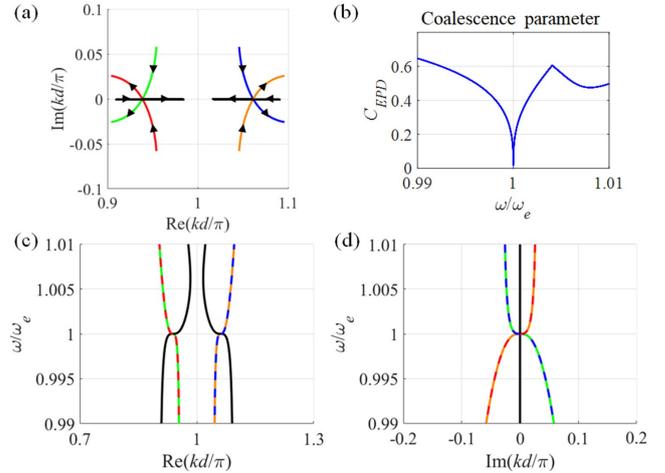

Fig. 5. Case $C$: The description is as in Fig. 3 but plots are for Case $C$.

These three different solutions show that our design to achieve the third order modal degeneracy in the three-way CTL is flexible, and the parameters of interest can be tuned around some initial practical values based on the application.

### D. Engineering of the Dispersion Diagram

One of the interesting features of the designs that we propose, which exhibit a third order modal degeneracy around a desired frequency, is that the slope of the dispersion diagram can be tuned easily by altering one or more design parameters of the unit cell. As a result, we can have a slightly positive local slope (small positive group velocity) or a slightly negative local slope (small negative group velocity) in proximity of the EPD, rather than the ideal case of zero slope. In Fig. 6(a), we show how the slope of the dispersion diagram for Case $A$, can be engineered to be positive or negative in the vicinity of the EPD by simply adjusting the value of the $R = 1/G$, for both the gain and radiation loss elements while still maintaining $GT$ symmetry in the system. We observe a slightly positive slope for slightly lower values of $R$ ($R = 5.27$ Ω) than the EPD one of $R = 7.15$ Ω, shown in solid blue. We observe a slightly negative slope for $R = 9.11$ Ω, i.e., slightly higher than the EPD one, shown in solid red. The case with $R = 7.15$ Ω that leads to the ideal third order EPD for case $A$ with zero slope is shown in solid black in Fig 6(a). In this figure, we only show the branches with purely real wavenumber, i.e., those with complex-valued $k$ are not shown for simplicity.

Another method to alter the slope of the dispersion diagram in the vicinity of the third order degeneracy is by tuning the height of the serpentine microstrip ($h$) as shown in Fig. 6(b). Note that by just altering $h$, the structure remains $GT$ symmetric. As observed from the results of Fig. 6(b), by slightly lowering the height ($h = 4.748$ mm) we achieve slightly positive slope (shown in solid blue) for the dispersion diagram of the case $A$ where the ideal EPD with zero slope occurs for $h = 5.348$ mm. Instead, by slightly increasing the height ($h = 5.948$ mm), we can also achieve a slightly negative slope (shown in solid red).

The results of these dispersion engineering examples demonstrate the flexibility of the proposed design for specific applications where the group velocity can be tuned by varying the design parameters about their nominal values. Increasing the slope of the dispersion diagram around the EPD frequency to reach a positive group velocity will potentially increase the bandwidth of the resonance peak associated to EPD, which will be desirable for reaching higher bandwidth-gain products in amplifier applications [55]. Alternatively, decreasing the slope to negative values in the dispersion diagram around EPD results in higher $Q$-factors for the EPD resonance peak [55], which may be beneficial for oscillator applications.

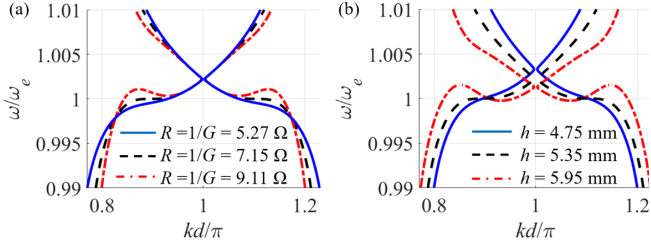

Fig. 6. Engineering of the dispersion diagram of the mode with purely real $k$ to exhibit different group velocities (different slopes) around the EPD frequency. (a) By tuning the value of the $R$ (or $G$) elements for Case $A$ we observe a slightly positive slope for $R = 5.27\Omega$ and a slightly negative slope for $R = 9.11\Omega$, whereas the ideal case with zero slope has $R = 7.15\Omega$. (b) By tuning the value of the $h$ (serpentine height) for Case $A$ we observe a slightly positive slope for $h = 4.748$ mm and a slightly negative slope for $h = 5.948$ mm, whereas the ideal case with zero slope has $h = 5.948$ mm. For all the graphs shown above only the purely real branches of the dispersion diagram are plotted and we have assumed the structure to be lossless.

### E. Power Analysis Based on Modes Around EPD

Exactly at the EPD ($\omega = \omega_e$), where three modes coalesce in their wavenumber $k_1 = k_2 = k_3 = k_e$, with $k_e$ purely real (Im($k_e$)=0), the eigenwaves propagating from unit cell to unit cell do not exhibit exponentially growing or decaying behavior. We checked this by using the single degenerate eigenvector as input state-vector in a semi-infinite structure i.e., $\mathbf{\Psi}(z = 0) = \mathbf{\Psi}_e$. This investigation of the power flow in the semi-infinite long periodic structure shows that at the EPD the power over $G$ and $-G$ is balanced, meaning they both have equal powers that cancel one another (i.e., $P_{-G} = -P_G$), as is discussed later in this section.

The obtained dispersion diagrams, shown in Fig. 3, 4 and 5, for the proposed $GT$-symmetric structures show that, at frequencies slightly lower or higher than the EPD frequency, the three modes are slightly perturbed from the EPD, and they are no longer coalescing. These three modes have one mode that has a purely real wavenumber (black curves in Fig. 3, 4 and 5), $k_1$, and the other two modes have wavenumbers that are complex conjugates of each other (red and green curves in Fig. 3, 4 and 5), $k_2 = k_3^*$. Another set of simulations were performed for the same proposed structures but with asymmetric gain and loss (broken $GT$ symmetry) and we found that the dispersion diagram did not exhibit the prementioned conjugate property for the wavenumbers $k_2 = k_3^*$.

Based on the prementioned conjugate property of the wavenumbers, the proposed $GT$-symmetric structures have two modal complex wavenumbers with Im($k_2$) = $-$Im($k_3$) which means that one mode is growing whereas the other one is decaying mode along $z$. One (red curve) of those two modal complex wavenumbers has Im ($k_2$) < 0 for $\omega < \omega_e$ whereas it has Im ($k_2$) >0 for $\omega > \omega_e$. For the purely real mode, $k_1$, there is no growing/decaying behavior in the signal and the power is balanced.

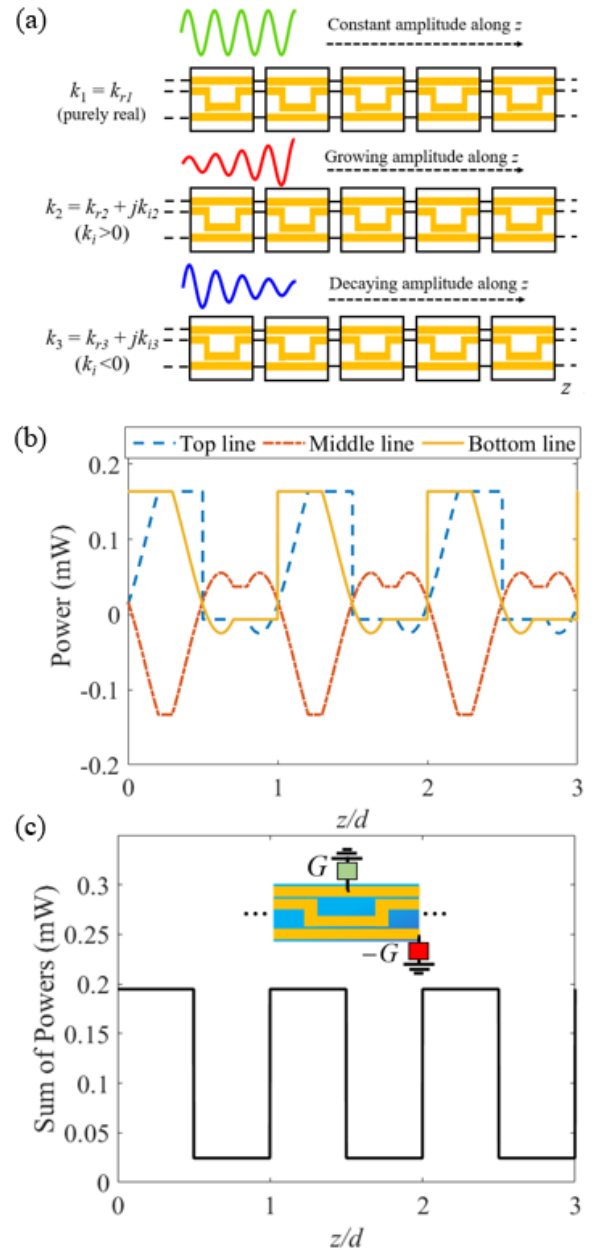

Fig. 7. (a) Graphical representation of how the signal/power propagates along the periodic structure in the +$z$ direction, for three different wavenumbers: purely real $k_1$, complex wavenumber with positive imaginary part ($k_2$), and with negative imaginary part ($k_3$). (b) Power (in mW) over the three lines of the semi-infinite long periodic structure plotted versus $z$ evaluated exactly at EPD frequency for the design of Case $A$, for the EPD in the region 0 < $k_e d$ < $\pi$. This plot exhibits how the power moves over the structure for the specific degenerate eigenmode on each line of the three-way waveguide. (c) Summation of the powers of the three lines shown in part (b), plotted versus $z$ for the semi-infinite periodic circuit. The jumps in the power are associated to the +$G$ and −$G$ contributions.

The mode with Im ($k$) > 0 has a growing behavior in the signal over the unit cells moving along z. For this case, there is more power provided by −$G$ (gain) than the power consumed by +$G$ (loss). Thus, the total power carried by this mode exiting a unit cell to the right is higher than the power entering the unit cell from the left. On the other hand, the mode with a negative imaginary part, has a decaying behavior, and in this case, there is more power consumed by +$G$ (loss) than the one provided by −$G$ (gain). A graphical summary of this investigation is presented in Fig. 7(a) where, for each of the three modes associated to a perturbed third order EPD, the propagating, growing, and decaying modes are shown. At the EPD the three modes coalesce to form one degenerate mode with a purely real $k = k_e$.

To better understand the power distribution inside the unit cells, we consider Case *A*: in Fig. 7(b) the power over the three lines of the semi-infinitely long periodic structure is plotted versus normalized *z*, evaluated exactly at EPD frequency and wavenumber such that $0 < k_e d < \pi$. The plot was obtained by assuming an input state vector at $z = 0$ is the EPD degenerate eigenvector, $\Psi(z = 0) = \Psi_e$, associated with positive value $k_e$ with $0 < k_e < \pi/d$, and the degenerate state eigenvector has been normalized such that $||\Psi_e|| = \Psi_e^T \Psi_e^* = 2.31 \text{ v}^2$. This eigenvector excites voltages and currents on each of the three TLs. The other degenerate mode with $-k_e$ has a different eigenvector, therefore the one used in this simulation, $\Psi_e$, excites only the three degenerate modes with positive $k_e$.

In Fig. 7(c), the total power (summation of the powers flowing in the three lines of the circuit, top, middle, and bottom) is plotted versus normalized *z*. We observe two different jumps in the power in each period that are associated to the power dissipation and contribution of the $+G$ and $-G$ lumped elements to the circuit, respectively. Since this power flow is evaluated exactly at the EPD condition, where the three coalesced wavenumbers are purely real, the power entering each unit cell from the left is the same as the one exiting to the right. Therefore, at the EPD the power over $+G$ and $-G$ is balanced, with $P_{-G} = -P_{+G}$. The conservation of the power is also verified in our numerical simulations by directly calculating these two quantities (the plot in Fig. 7(c) is obtained by summing the three powers) where we see that the total power carried by this degenerate mode exiting any unit cell to the right is equal to the power entering it from the left.

We have selected Case *A* for demonstration purposes in Fig. 7(b) and Fig. 7(c), but the general concept of the power analysis provided here is analogous with the other two cases featuring third order EPDs. This study provides us with some physical insight into how different modes behave and how the signal/power is propagating throughout the structure. In the following section, we will provide more investigation of the powers and gain for a finite-length and terminated periodic structure.

### III. FINITE-LENGTH STRUCTURE PROPERTIES

As discussed earlier, devices featuring EPDs may exhibit special properties and enhanced characteristics which make them potential candidates for applications. To provide an example application of the regime presented in this paper, we consider a finite-length three-way waveguide constructed by cascading the proposed *GT*-symmetric unit cells and adding proper excitation and terminations to make a distributed amplifier, with the $-G$ as distributed gain and with the $+G$ elements as radiative loads (modeling antennas). In this section, we first provide an investigation of the resonance behavior and stability analysis of such finite-length three-way waveguide and then show the amplification at the EPD frequency.

#### A. Resonance Behavior and Stability Analysis

We consider the three-way waveguide structure of Case *A* in Fig. 2(a), consisting of *N* cascaded unit cells as depicted in Fig. 8(a). We have omitted the right-most $-G$ element, as shown, to make the terminated structure symmetric and help to improve stability. We excite the middle line of the three CTLs with terminations of $Z_s = Z_L = 50$ Ω. For the terminals of the bottom line, we are assuming $Z_y = 50$ Ω, and for the top line we are assuming short circuit terminations ($Z_x = 0$ Ω) as shown in Fig. 8(a). We have selected this loading scenario based on the stability and gain performance of the three-way structure. First, to check the stability and the resonance behavior, we check the *S* parameters. Based on [67], for two port networks, oscillations are possible when either the input or output port present a negative resistance, which occurs when $|S_{11}|>1$ or $|S_{22}|>1$ in our structure setup, treated as a two port network (because of symmetry, $S_{11} = S_{22}$). To check stability, we need to evaluate $S_{11}$. For the design of Case *A*, the results for the $S_{11}$ and $S_{21}$ parameters, assuming *N*=8 unit cells and lossless structure (besides the lumped elements), is provided in Fig. 8(b) over a wide frequency range. The structure is stable based on the $S_{11}$ response shown in Fig. 8(b). For other configurations or loading scenarios, stability could also be reached by using impedance matching circuits (filters). The $S_{21}$ parameter plotted in Fig 8(b) versus frequency shows a sharp resonance peak denoted by $\omega_r$ associated to the third order EPD frequency of 2GHz.

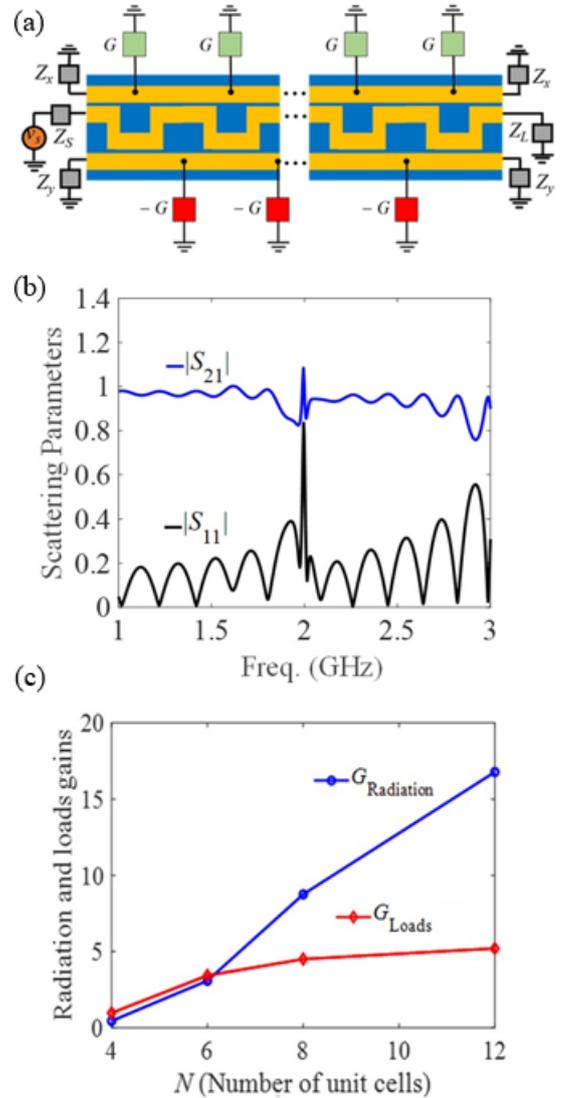

Fig. 8. (a) Finite-length three-way structure with period *d* made by cascading *N* unit cells with the total length of $L = Nd$. (b) Plot of $|S_{11}|$ and $|S_{21}|$ for the finite length structure of Case *A* with *N*=8 unit cells and substrate with tan($\delta$)=0, around the EPD frequency, $f_e$= 2 GHz, where we observe that $|S_{11}|<1$ and hence unconditionally stability. (c) Radiation gain and loads gain (both in linear scale) versus the length *N* of the finite structure of Case *A*, evaluated at the strong-peak frequency nearest to the EPD. For all results shown above we have assumed $Z_y = Z_s = Z_L = 50$Ω for the bottom and middle lines, and two short circuits for the top line ($Z_x = 0$ Ω).

#### B. Gain Evaluation

To evaluate the behavior of the proposed distributed amplifier, we consider the power delivered to the loads ($Z_L$ and $Z_y$) as well as the power delivered to all the *N* passive elements ($+G$) for different structure lengths. We calculate the load power gain ($G_{\text{Loads}}$) and radiation power gain ($G_{\text{Radiation}}$). In our analysis, the load gain is defined as $G_{\text{Loads}} = P_{\text{Loads-total}} / P_{\text{in}}$ in which $P_{\text{Loads-total}}$ is summation of the power over $Z_L$ and the two $Z_y$, and $P_{\text{in}}$ is the input power at the second (middle) line input. The radiation gain is defined as $G_{\text{Radiation}} = P_{\text{Radiation-total}} / P_{\text{in}}$, in which $P_{\text{Radiation-total}}$ is the summation of the powers delivered to the $+G$ elements and $P_{\text{in}}$ is the input power at the second (middle) line input. All the other parameters are the same as previously discussed for case $A$.

In Fig 8(c), the radiation gain and load gain are plotted versus the length of the finite-length structure ($N$) at the strong-peak resonance frequency nearest to the third order EPD in Case $A$, still assuming absence of lossless in the substrate, $\tan(\delta)=0$, and in the metals. We observe high values of radiation gain, significantly larger than the load gain. These results are based upon the terminations of $Z_y = Z_s = Z_L = 50$ Ω for the bottom and middle lines and two short circuits for the top line ($Z_x = 0$ Ω), similar to the previous section. The result of Fig. 8(c) shows that, while being stable, for the case of $N = 8$, we reach a radiation gain of $G_{\text{Radiation}} = 8.8$ for the passive radiating elements with $+G$, while the load gain has a lower value of $G_L = 4.5$ at the EPD resonance frequency. The radiation gain increases significantly by increasing the radiator's length, which makes the proposed structure a potential scheme for distributed amplifier applications.

## IV. CONCLUSION

We have reported the existence of third order EPDs with real valued wavenumbers in three-way waveguides with a $GT$ symmetry. At the EPD, three eigenmodes coalesce at a desired frequency and purely real wavenumber. Besides having a real valued wavenumber in the presence of gain and loss, there is also one branch (solid black, Figs. 1,3,4, and 5) of the dispersion diagram which has purely real wavenumbers.

We have provided two different waveguide configurations, demonstrated how the group velocity of the mode with purely real wavenumber can be slightly altered by tuning the physical parameters, which may be beneficial for various applications. A potential scheme using this third order EPD could be in high-gain distributed amplifiers with distributed power extraction. Indeed, the simultaneous presence of distributed gain and losses (modeling radiation conductances) and the same slope sign of the propagating-wavenumber branch (black curves in Figs. 1,3,4, and 5) at frequencies below and above the EPD frequency, paves the way to a new set of applications of EPDs in high power radiating "apertures". We have briefly discussed such an application and provided the radiation gain analysis for finite-length array of antennas, where each antenna is represented by a lumped "radiation resistance". The fundamental idea here presented is not limited to the specific design shown in this paper but can be potentially applied to a variety of periodic waveguide structures implemented in different technologies, including EPD lasers with distributed power extraction.

Importantly, the kind of third order EPD studied in this paper is exhibited in the presence of periodic gain and antennas (loss), so arrays of this kind can radiate high power if gain and loss are designed to be large. This is very different from the concept of an SIP (i.e., frozen mode) in a lossless/gainless waveguide, where distributed gain was then introduced as in [55]; in that case, the SIP is increasingly destroyed when higher and higher gain is introduced in each unit cell, whereas the 3$^{\text{rd}}$ order EPD in this paper is fully maintained even with large gain elements if properly designed, enabling very high power applications of EPDs. Examples of a second order EPD in waveguiding structures that exists while high power is continuously extracted along the waveguide are provided in the oscillator concept shown in [48], and in the backward oscillator concept presented in Ref. [47,60,61] leading to high power and high efficiency. Analogously, the third order EPD shown in this paper can be exploited for high power radiating oscillators, lasers with distributed power extraction, and distributed amplifiers with distributed power extraction.

### ACKNOWLEDGMENT

This material is based upon work supported by the National Science Foundation under award NSF ECCS-1711975 and by the Air Force Office of Scientific Research under award number FA9550-18-1-0355.

### APPENDIX A: PARAMETERS USED IN SIMULATIONS

In our simulations, we considered a periodic coupled three-way waveguide composed of unit cells, each made of three coupled TLs as in Fig. 2. For all the designs discussed in this paper, the microstrip line widths are fixed to have $w = 5$ mm (i.e., with 50 Ω characteristic impedance) and $s = 0.5$ mm for the distancing between the lines. The substrate is assumed to have a relative dielectric constant of 2.2, loss tangent of 0 (lossless dielectric), and thickness of $h_s = 1.575$ mm. Metal layers are assumed to be lossless as well.

Case $A$: The tuned unit-cell parameters that led to an EPD were found to have conductance values of $G = 0.1398$ S (or equivalently $R = 1/G = 7.15$ Ω), serpentine height of $h = 5.35$ mm, and period of $d = 54.15$ mm.

Case $B$: For this case, the tuned unit-cell parameters have a conductance value of $G = 0.105$ S (or equivalently $R = 1/G = 9.5$ Ω), serpentine height of $h = 6.36$ mm, and period of $d = 46.3$ mm.

Case $C$: For this case, the tuned parameters have a conductance value of $G = 0.0099$ S (or equivalently $R = 1/G = 100.55$ Ω), serpentine height of $h = 1.07$ mm, and period of $d = 48.08$ mm.

### APPENDIX B: TRANSFER MATRIX FORMALISM

#### A. Transfer Matrices for CTLs

In order to construct the transfer matrix and tune the physical unit cell dimensions to acquire a third order EPD, we have divided the unit cell of the three-way microstrip waveguide into smaller segments as shown in Fig. 1 and modeled each segment to obtain the unit-cell transfer matrix. We built the T-matrix of each segment using TL analytic formulas based on quasistatic models in [68,69].

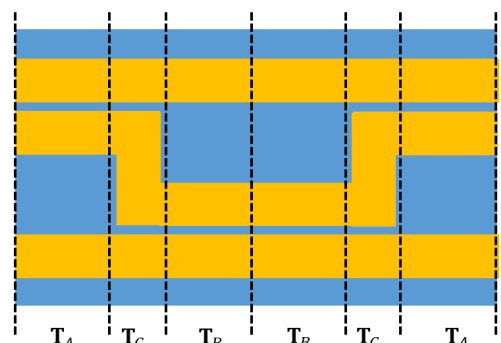

Fig. 9. Unit cell of the 3-way periodic microstrip structure over a grounded substrate used to obtain the EPD, divided into smaller segments to construct the corresponding transfer function for each segment in terms. The total transfer matrix of the unit cell is then derived by multiplying the segment transfer matrices.

The transfer matrices of each smaller segment of the unit cell shown in Fig. 1(a) are expressed and calculated in terms of the parameters of the unit cell of the system (length, width, height, separation). Finally, the transfer matrix for the whole unit cell (without the added conductances and gain elements) is obtained by the product of the transfer matrices for each smaller segment of three CTLs inside the unit cell as

$$\underline{\mathbf{T}}_U = \underline{\mathbf{T}}_A \underline{\mathbf{T}}_C \underline{\mathbf{T}}_B \underline{\mathbf{T}}_B \underline{\mathbf{T}}_C \underline{\mathbf{T}}_A. \tag{A1}$$

### B. Transfer Matrix for Lumped Conductances

We find the transfer matrices for the added conductance and gain lumped elements in each unit cell. For the first admittance added on the top line with the value of $-G$ (active device) we have

$$\underline{\mathbf{T}}_{+G} = \begin{pmatrix} 1 & 0 & 0 & & \\ GZ_0 & 1 & 0 & & \underline{\mathbf{0}}_{3\times3} \\ 0 & 0 & 1 & & \\ & \underline{\mathbf{0}}_{3\times3} & & & \underline{\mathbf{I}}_{3\times3} \end{pmatrix}, \tag{A2}$$

where $G$ is the conductance value (assumed positive) of the active gain device. For the second admittance added on the bottom line with the value of $G$ (passive device) we have

$$\underline{\mathbf{T}}_{-G} = \begin{pmatrix} \underline{\mathbf{I}}_{3\times3} & & & \underline{\mathbf{0}}_{3\times3} \\ & 1 & 0 & 0 \\ \underline{\mathbf{0}}_{3\times3} & 0 & 1 & 0 \\ & 0 & -GZ_0 & 1 \end{pmatrix}. \tag{A3}$$

Therefore, the total transfer matrix for the unit cell of Case $A$ shown in Fig. 2(a), including the added lumped radiation conductance and gain device, is calculated as

$$\underline{\mathbf{T}}_U = \underline{\mathbf{T}}_{-G} \underline{\mathbf{T}}_A \underline{\mathbf{T}}_C \underline{\mathbf{T}}_B \underline{\mathbf{T}}_{+G} \underline{\mathbf{T}}_B \underline{\mathbf{T}}_C \underline{\mathbf{T}}_A. \tag{A4}$$

The total transfer matrix for the unit cell of Case $B$ shown in Fig. 2(b), including the added lumped radiation conductance and gain device, is calculated as

$$\underline{\mathbf{T}}_U = \underline{\mathbf{T}}_G \underline{\mathbf{T}}_A \underline{\mathbf{T}}_C \underline{\mathbf{T}}_B \underline{\mathbf{T}}_{-G} \underline{\mathbf{T}}_B \underline{\mathbf{T}}_C \underline{\mathbf{T}}_A. \tag{A5}$$